\documentclass{article}
\usepackage{amssymb,slashed,framed}

\usepackage{graphicx}
\usepackage{amsmath,mathrsfs}

\begin{document}

\title{A Note on the Relativistic Transformation Properties of Quantum Stochastic
Calculus}% Force line breaks with \\
%\thanks{Or quantum time travel}%

\author{John E. Gough\\
 Department of Physics, Aberystwyth University, \\
 Wales, UK, SY23 3QR.}%
 
\date{\today}% It is always \today, today,

\maketitle

\begin{abstract}
We give a simple argument to derive the transformation of quantum stochastic calculus formalism between inertial observers, and derive the quantum open system dynamics for a system moving in a vacuum (more generally coherent) quantum field under the usual Markov approximation. We argue that for uniformly accelerated open systems, however, the formalism must breakdown as we move from a Fock representation over the algebra of field observables over all Minkowski space to the restriction to the algebra of observables over a Rindler wedge. This leads to quantum noise having a unitarily inequivalent non-Fock representation - in particular, the latter is a thermal representation at the Unruh temperature. The unitary inequivalence ultimately being a consequence of the underlying flat noise spectrum approximation for the fundamental quantum stochastic processes. We derive the quantum stochastic limit for a uniformly accelerated (two-level) detector and establish an open systems description of the relaxation to thermal equilibrium at the Unruh temperature.
\end{abstract}

\section{Introduction}
The quantum stochastic calculus was introduced by Hudson and Parthasarathy \cite{HP84} and by Gardiner and Collett \cite{GC} as a technique for describing open quantum systems. Typically, the quantum system is considered to be at rest with the quantum noise entering as quantum input processes (quantum white noises). The question of a relativistically covariant formulation was addressed and solved by Frigerio and Ruzzier \cite{Frigerio_Ruzzier}. Their construction makes extensive use of unitary representations of the Poincar\'{e} group on the one-particle space for the input noise, however, we present a shorter and more direct argument based on quantum white noises.

In Section \ref{sec:RQSC}, we give a direct argument which re-derives the transformation rules for quantum Ito table for quantum stochastic calculus when we go from one inertial frame to another.

One envisages a quantum mechanical system moving along a world line. Here we adopt a semi-classical view that the position of the
system is resolved only on length scales much larger than its de Broglie wavelength: in particular, its quantum mechanical nature is described by internal degrees of freedom only (excluding specifically its position and momentum observables) for which the underlying Hilbert space is $\mathfrak{h}_{0}$. In Section \ref{sec:OpenSystem}, we consider the quantum open systems models where we have a unitary evolution for the system and background field. We also consider coherent state fields.  

Finally, in Section \ref{sec:Unruh} we consider the Unruh effect for the case of uniformly accelerated systems. Here we disagree with the claim of \cite{Frigerio_Ruzzier} that the transformation rules apply without modification to arbitrary moving observers. In fact, the non-Fock nature of the noise for uniformly accelerated observers means that the gauge process is not well defined. We shall however, derive and solve the master equation in this regime.

\section{Relativistic Transformations of Quantum  Stochastic Calculus}
\label{sec:RQSC}
Let $K$ and $K^{\prime }$ be inertial frames in standard configuration with spacetime coordinates related by the Lorentz transformation
\begin{eqnarray}
t^{\prime }=\gamma \left( u\right) \left( t-\frac{u}{c^{2}}x\right)
,
\qquad x^{\prime }=\gamma \left( u\right) \left( x-ut\right) ,
\qquad  y^{\prime }=y,\qquad z^{\prime }=z
,
\end{eqnarray}
where $\gamma \left( u\right) =1/\sqrt{1-\frac{u^{2}}{c^{2}}}$. (The
interpretation is that $K^{\prime }$ moves along the $x$-axis of $K$ with
velocity $u$.) In the following, we will ignore the other two space dimensions.

A system is assumed to follow a world line given by $x=vt$ in $K$: that is, it moves with velocity $v$ along the $x$-axis in $K$, and therefore is seen by $K^{\prime }$ to move along the $x^{\prime }$-axis with velocity
\begin{eqnarray}
v^{\prime }=\frac{v-u}{1-\frac{uv}{c^{2}}} .
\end{eqnarray}
If we consider two events along the world line of the system, then the proper time $\Delta \tau $ that elapses in the rest frame of the system is related to the times $\Delta t$ and $\Delta t^{\prime }$ measured in $K$ and  $K^{\prime }$, respectively, by 
\begin{eqnarray}
\Delta \tau =\sqrt{1-\frac{v^{2}}{c^{2}}}\Delta t=\sqrt{1-\frac{v^{\prime 2}%
}{c^{2}}}\Delta t^{\prime }.
\end{eqnarray}
In other words, $\Delta \tau =\frac{1}{\gamma \left( v\right) }\Delta t=\frac{1}{\gamma
\left( v^{\prime }\right) }\Delta t^{\prime }$. It follows that
\begin{eqnarray}
\Delta t^{\prime }=\zeta \left( u,v\right) \Delta t
\end{eqnarray}
where $\zeta \left( u,v\right) =\gamma \left( v^{\prime }\right) /\gamma
\left( v\right) $ and after a little algebra, one obtains
\begin{eqnarray}
\zeta \left( u,v\right) =\gamma \left( u\right) \left( 1-\frac{uv}{c^{2}}%
\right) .
\label{eq:zeta}
\end{eqnarray}

Note that if $u\equiv v$, then $K^{\prime }$ is co-moving with the system and
so $\Delta t^{\prime }\equiv \Delta \tau $. Here $\zeta \left( v,v\right)
\equiv 1/\gamma \left( v\right) $ which is the correct time dilation factor
between $\Delta \tau $ and $\Delta t$.

\subsection{Quantum White Noise in the Rest Frame}

We construct the quantum stochastic calculus from quantum white noises. In the rest frame of the system, we introduce annihilation and creation operators $\bar{b}\left( \tau \right) $ and $\bar{b}^{\ast }\left( \tau \right) $ satisfying the singular commutation relations
\begin{eqnarray}
\left[ \bar{b}\left( \tau \right) ,\bar{b}^{\ast }\left( \tau ^{\prime
}\right) \right] =\delta \left( \tau -\tau ^{\prime }\right) .
\end{eqnarray}
The quantum stochastic processes may then be defined as
\begin{eqnarray}
\textbf{(The Gauge Process)} \qquad 
\bar{\Lambda}\left( \tau \right) &=&\int_{0}^{\tau }\bar{b}^{\ast }\left(
\sigma \right) \bar{b}\left( \sigma \right) d\sigma ,\\
\textbf{(The Annihilation Process)} \qquad \bar{B}\left( \tau
\right) &=& \int_{0}^{\tau }\bar{b}\left( \sigma \right) d\sigma ,\\
\textbf{(The Creation Process)} \qquad 
\bar{B}^{\ast
}\left( \tau \right) &=& \int_{0}^{\tau }\bar{b}^{\ast }\left( \sigma \right)
d\sigma .
\end{eqnarray}
This leads to the quantum It\={o} table \cite{HP84}
\begin{eqnarray}
\begin{array}{ll}
d\bar{\Lambda}\left( \tau \right) d\bar{\Lambda}\left( \tau \right)  = d\bar{\Lambda}\left( \tau \right) , & 
d\bar{B}\left( \tau \right) d\bar{\Lambda}%
\left( \tau \right) =d\bar{B}\left( \tau \right)   \\
d\bar{\Lambda}\left( \tau \right) d\bar{B}^{\ast }\left( \tau \right)  =d \bar{B}\left( \tau \right) , & d\bar{B}\left( \tau \right) d\bar{B}^{\ast }\left( \tau \right) = d\tau .
\end{array}
\label{eq:qITOtable}
\end{eqnarray}

\subsection{Referring to a Moving Frame}
In order to work out the corresponding processes in the inertial frame $K$,
we observe that the process $\bar{\Lambda}\left( \tau \right) $ should count
noise quanta along the world line from proper time 0 up to proper time $\tau $%
. This must, of course, be observer independent so we identify $ \bar \Lambda (\tau )$ with the
process $\Lambda \left( t\right) =\int_{0}^{t}b^{\ast }\left( s\right)
b\left( s\right) ds$. Therefore, $\bar{b}^{\ast }\left( \tau \right) \bar{b}%
\left( \tau \right) d\tau \equiv b^{\ast }\left( t\right) b\left( t\right) dt
$, and using $dt=\gamma \left( v\right) d\tau $, we deduce that
\begin{eqnarray}
b\left( t\right)  \equiv \frac{1}{\sqrt{\gamma \left( v\right) }}\, \bar{b}\left( \tau
\right) .
\end{eqnarray}

Therefore, $dB\left( t\right) =b\left( t\right) dt=\sqrt{\gamma \left( v\right) }\, \bar{b}
\left( \tau \right) d\tau \equiv \sqrt{\gamma \left( v\right) }d\bar{B}\left( \tau
\right) $. The full set is
\begin{eqnarray}
\begin{array}{ll}
d \Lambda (t) = d \bar \Lambda (\tau ) , &  dB\left( t\right) = \sqrt{\gamma \left( v\right) }\, d\bar{B}\left( \tau
\right) 
 ,  \\
dB^\ast \left( t\right) = \sqrt{\gamma \left( v\right) }\, d\bar{B}^\ast \left( \tau
\right) ,& dt = \gamma (v) \, d \tau .
\end{array}
\label{eq:trans_gamma}
\end{eqnarray}
The corresponding Ito table in inertial frame $K$ is readily deduced to be
\begin{eqnarray}
\begin{array}{ll}
d\Lambda \left( t\right) d\Lambda \left( t\right)  = d\Lambda \left(
t\right) , & dB\left( t\right) d\Lambda \left( t\right) =dB,  \\
d\Lambda \left( t\right) dB^{\ast }\left( t\right)  = dB^{\ast }\left(
t\right) , & dB\left( t\right) dB^{\ast }\left( t\right) =dt.
\end{array}
\label{eq:qITOtable2}
\end{eqnarray}

For instance, $dB\left( t\right) d\Lambda \left( t\right) =\sqrt{\gamma
\left( v\right) }d\bar{B}\left( \tau \right) d\bar{\Lambda}\left( \tau
\right) =\sqrt{\gamma \left( v\right) }d\bar{B}\left( \tau \right) =dB\left(
t\right) $, etc.

\subsection{Transformation Between Inertial Frames}
The transformation law from $K$ to $K^{\prime }$ is then easily obtained by
noting that 
\begin{eqnarray}
    b^{\prime }\left( t^{\prime }\right) =\sqrt{\gamma \left(
v\right) /\gamma \left( v^{\prime }\right) } \, b\left( t\right) =\frac{1}{\sqrt{\zeta \left( u,v\right) }} \, b\left( t\right) .
\end{eqnarray}
This yields the transformation
\begin{eqnarray}
\begin{array}{ll}
d\Lambda ^{\prime } =d\Lambda , & dB^{\prime } =\sqrt{\zeta \left( u,v\right) } \,
dB,  \\
dB^{\prime \ast } =\sqrt{\zeta \left( u,v\right) }\, dB^{\ast }, &
dt^{\prime
} = \zeta \left( u,v\right) dt.
\end{array}
\label{eq:FR_trans}
\end{eqnarray}
 In \cite{Frigerio_Ruzzier}, the calculations are presented in terms of hyperbolic angles $\alpha,\beta$ where $\tanh \alpha = - \frac{u}{c} $ and $\tanh \beta = \frac{v}{c}$. They obtain the transformations (\ref{eq:FR_trans}) with $\zeta = \cosh \alpha + \beta \sinh \alpha$, however, this by inspection agrees with our (\ref{eq:zeta}).

\subsection{Fermi Noise}
 The above arguments can be applied to derive the transformation rues for fermionic fields without additional work. The Fermi \cite{HP_fermi} version can be constructed from anti-commuting processes satisfying $\{ a(t) , a^\ast (s) \} = \delta (t-s)$.
 It follows that the Fermi processes will again transform in the same way, i.e., equations (\ref{eq:trans_gamma}) and (\ref{eq:FR_trans}), as their Bose counterparts.
 
\section{Unitary Evolutions}
\label{sec:OpenSystem}
We recall that the standard form of a unitary quantum stochastic differential equation describing our system (with underlying Hilbert space $\mathfrak{h}_0$) interacting with the noise is \cite{HP84}
\begin{eqnarray}
    d \bar  U(\tau ) =  d\bar{G} (\tau ) \, \bar U(\tau ) , \qquad \bar U (0) = I,
    \label{eq:qsde}
\end{eqnarray}
where
\begin{eqnarray}
    d \bar G (\tau ) = (\bar S-I) \otimes d \bar \Lambda (\tau )
    +\bar L \otimes d \bar B^\ast (\tau ) - \bar L^\ast \bar S \otimes d\bar B (\tau) 
    - ( \frac{1}{2} \bar  L^\ast \bar L +i \bar  H) \otimes d \tau .
    \label{eq:generator}
\end{eqnarray}
 The operators $\bar  S, \bar L, \bar  H$ on $\mathfrak{h}_0$ are assumed to be unitary, bounded, and bounded self-adjoint, respectively.

\subsection{Moving frame QSDE}
 The picture for reference frame $K$ will be described by the same equations as (\ref{eq:qsde}) and (\ref{eq:generator}) with the bars removed from the noise operators, the proper time $\tau$ replaced by $t$: $ d  U(t ) =  dG (t) \, U(t) , \, U (0) = I$,
 \begin{eqnarray}
    d G (t ) = (S-I) \otimes d  \Lambda (t )
    + L \otimes d  B^\ast (t) \nonumber - L^\ast  S \otimes d B (t) 
    - ( \frac{1}{2}   L^\ast  L +i  H) \otimes d t
    \label{eq:generator_K}
\end{eqnarray}
 and the new coefficient operators are as follows
 \begin{eqnarray}
     S = \bar S, \quad L = \frac{1}{\sqrt{\gamma (v) }} \, \bar L , \quad
     H = \frac{1}{ \gamma (v) } \, \bar H .
 \end{eqnarray}
A quantum dynamical semi-group $(\bar \Phi_ \tau )_{\tau \ge 0}$ on operators of $\mathfrak{h}_0$ is defined by
\begin{eqnarray}
    \langle \psi_1 | \bar \Phi_\tau (X) | \psi_2 \rangle = 
    \langle \psi_1 \otimes \mathrm{vac} | \bar U ^\ast(\tau )
    (X \otimes I ) \bar U (\tau ) |  \psi_2 \otimes \mathrm{vac} \rangle.
    \label{eq:semigroup}
\end{eqnarray}
Its Lindblad generator \cite{Lindblad} is
\begin{eqnarray}
    \bar{ \mathcal{L}} (X) = 
    -i [ X, \bar H ] + \frac{1}{2} [ \bar L^\ast , X ]\bar L
    + \frac{1}{2} \bar L^\ast [X, \bar L] .
\end{eqnarray}
We therefore deduce that the forms in $K$ and $K^\prime$ will be
\begin{eqnarray}
    \mathcal{L} (\cdot ) = 
    \zeta (u,v) \, \mathcal{L}^\prime (\cdot )=\frac{1}{\gamma (v)} \bar{\mathcal{L}}  (\cdot ) .
    \label{eq:scale_Lindblad}
\end{eqnarray}
This agrees with the expectation that $ \bar{\mathcal{L}} \, d\tau = \mathcal{L} \, dt = \mathcal{L}^\prime \, dt^\prime $.
See \cite{SLH,Series_Product,CKS}.

\subsection{Coherent states}
In general we can consider the incoming quantum noise to be in a coherent state $| \bar \alpha \rangle$ where $\bar \alpha (\tau ) $ gives the complex amplitude in the rest frame at proper time $\tau$. We obtain this state by displacing the Fock vacuum by
\begin{eqnarray}
    | \bar \alpha \rangle = e^{- \| \bar \alpha \|^2} e^{\int \bar \alpha (\tau ) \bar b ^\ast (\tau ) \, d\tau } | \mathrm{vac}  \rangle .
\end{eqnarray}
The normalization involving $\| \bar \alpha \|^2 = \int | \bar \alpha (\tau ) |^2 \, d \tau$ which is assumed finite.

In the inertial frame $K$, we will have the complex amplitude
\begin{eqnarray}
    \alpha (t) = \frac{1}{\sqrt{ \gamma (v)}} \bar \alpha (\tau ).
\end{eqnarray}
This follows immediately by identifying $\int \bar \alpha (\tau ) \bar b ^\ast (\tau ) \, d\tau $ with $\int \alpha (t) b^\ast (t) \, dt$.

We may replace the vacuum state in (\ref{eq:semigroup}) with the coherent state $| \bar \alpha \rangle$. The effect of this can be modeled by making the replacements
\begin{eqnarray}
    \bar L &\to& \bar L _{\bar \alpha } =\bar L + \bar S \, \bar \alpha (\tau ), \nonumber \\
    \bar H &\to& \bar H _{\bar \alpha } = \bar H + \frac{1}{2i} \big(
    \bar L^\ast \bar \alpha (\tau ) - \bar L \bar \alpha ^\ast (\tau ) \big).
\end{eqnarray}
and averaging in the vacuum state \cite{GK}. The instantaneous Lindblad generator is then
\begin{eqnarray}
    \bar{\mathcal{L}}_{\bar \alpha } (X) = \bar \alpha^\ast (\tau ) \big( \bar S^\ast X \bar S - X) \bar \alpha (\tau )
    + \bar \alpha^\ast (\tau ) \bar S^\ast [ X, \bar L]
    +[ \bar L^\ast , X ] \bar S \bar \alpha (\tau )
    + \bar{\mathcal{L}} (X) .
\end{eqnarray}
We may likewise construct the Lindblad generators $\mathcal{L}_\alpha$ and $\mathcal{L}^\prime_{\alpha^\prime}$ for the coherent state in $K$ and $K^\prime$, respectively, however, it is apparent that they scale in exactly the same was as in (\ref{eq:scale_Lindblad}).

\section{Uniformly Accelerated Systems}
\label{sec:Unruh}
In \cite{Frigerio_Ruzzier} the situation of rectilinear motion is treated (that is, the system is modeled in a laboratory that has a fixed inertial reference frame). However, the claim is made that the invariance of the quantum stochastic calculus should extend to arbitrary motions as well. The only complication is that the proper time now described by an integral expression.

We argue that this can only be true for situations where the acceleration is negligible. Specifically, the approach must breakdown if we consider uniformly accelerated systems as here we will additionally encounter the Unruh effect \cite{Unruh,Davies,Crispino} which predicts that the system will thermalize with inverse (Unruh) temperature
\begin{eqnarray}
    \beta = \frac{2 \pi c}{a \hbar} 
\end{eqnarray}
where $a$ is the constant proper acceleration.

Suppose that our system is a cavity mode $c$ with resonant frequency $\Omega$ then the thermalization of an oscillator mode is modeled by the quantum markovian evolution $dU(t) = dG(t) \, U(t)$ with
\begin{eqnarray}
    dG(t) = K\otimes dt + \sqrt{\gamma} c \otimes dA^\ast(t)- \sqrt{\gamma} c^\ast \otimes dA(t) 
\end{eqnarray}
where $\gamma >0$ is a damping constant, $K = - \big( \frac{1}{2} (2n+1 )\gamma + i\hbar \Omega \big) c^\ast c$ and the processes $A(t),A(t)^\ast$ are non-Fock quantum Wiener process satisfying the quantum Ito table \cite{HL85}
\begin{gather}
    dA(t) dA(t)^\ast = (n+1) \, dt, \quad
    dA(t)^\ast dA(t) = n \, dt, \nonumber \\
    dA(t)dA(t)=0=dA(t)^\ast d A(t)^\ast .
\end{gather}
The parameter $n>0$ and should be set to the average \textit{boson} occupation number for an energy $\hbar \Omega$ at inverse temperature $\beta$:
\begin{eqnarray}
    n \equiv \frac{1}{e^{\beta \hbar \Omega}-1}
    \label{eq:n_Unruh}
\end{eqnarray}
As is well known, the non-Fock processes may be obtained from the standard quantum Ito table (\ref{eq:qITOtable}) or (\ref{eq:qITOtable2}) according to the Araki-Woods type construction \cite{ArakiWoods}
\begin{eqnarray}
    A(t) = \sqrt{n+1} \, B_1(t)\otimes I + \sqrt{n}\, I \otimes B_2 (t)^\ast .
    \label{eq:AW}
\end{eqnarray}

Unfortunately, it is impossible to implement the transformation from the $B$-processes to the $A$-processes unitarily as the conditions of Shale's Theorem are not meet in this case, see for instance \cite{ParthQSC}. Notably, it is impossible to include an analogue of the counting process $\Lambda (t)$ into the non-Fock calculus.

At this stage, it is worth reviewing the status of the Unruh effect itself. It has been unequivocally established theoretically that an observer moving in a (zero-temperature) vacuum will experience will undergo a convergence to thermal equilibrium at the Unruh temperature. However, the existence of an Unruh radiation was first challenged in \cite{Grove} and later \cite{Raine}. The latter paper used standard markovian approximations from quantum optics (Wigner-Weisskopf) however this is not an essential objection: it was conclusively shown in the work of Ford and O'Connell \cite{FOC_Unruh}, in the framework of the quantum Langevin equation, that there is no actual Unruh radiation as such.

In \cite{Crispino}, equations (2.122) and (2.123) with $c=1$, it is shown that the relevant annihilators associated with a uniform accelerated observer (corresponding to the right/left-moving Rindler modes in the right-hand wedge) take the form of a Bogoliubov transformation
\begin{eqnarray}
    \hat a ^R_{ \omega , \mathbf{k}_\perp} &=& \frac{ \hat b_{- \omega , \mathbf{k}_\perp}+ e^{-\pi  \omega /a} \hat b ^\ast_{+\omega , - \mathbf{k}_\perp}}{\sqrt{1-e^{-2\pi  \omega /a}}}, \nonumber\\
    \hat a ^L_{ \omega , \mathbf{k}_\perp} &=& \frac{ \hat b_{+ \omega , \mathbf{k}_\perp}+ e^{-\pi  \omega /a} \hat b ^\ast_{-\omega , - \mathbf{k}_\perp}}{\sqrt{1-e^{-2\pi \omega /a}}}.
\end{eqnarray}
The modes are labeled by the frequency parameter $\omega >0$ and the transverse wavenumber $\mathbf{k}_\perp$.

The $b$ modes satisfy standard commutation relations $[\hat b_{\pm \omega , \mathbf{k}_\perp},\hat b_{\pm \omega^\prime , \mathbf{k}^\prime_\perp}^\ast] = \delta( \omega - \omega ') \delta (\mathbf{k}_\perp - \mathbf{k}_\perp^\prime)$ and the $\hat b_{\pm \omega , \mathbf{k}_\perp}$ annihilate the vacuum $|0 _M \rangle$ of Minkowski space.

In the following, we shall make some typical quantum optics approximation but do not make any pretense at rigor. One first ignores the transverse labels $\mathbf{k}_\perp$, then assumes that the field couples to the quadrature $c+c^\ast$ of the oscillator which, in the interaction picture, picks up phases $e^{\pm i \Omega \tau }$, then one assumes  the main contribution from the field comes from those modes $\omega $ close to $\Omega$ and make a rotating-wave approximation, finally one reintroduces a flat spectrum assumption for the modes leading to a white noise approximation. This motivates the limit relation
\begin{eqnarray}
     a^R (t) = \sqrt{n+1}\, b_- (t) \otimes I + \sqrt{n}\, I \otimes b_+ (t)^\ast , \nonumber \\
     a^L (t) = \sqrt{n+1}\, b_+ (t) \otimes I + \sqrt{n}\, I \otimes b_- (t)^\ast ,
     \label{eq:AWqwn}
\end{eqnarray}
with $n$ being the correct Unruh average number (\ref{eq:n_Unruh}). 

Note that $a^R$ and $a^L$ will be two \textit{commuting} quantum white noise processes: $[a^R (t) , a^L (s) ]= \sqrt{(n+1)n}\big( \delta(t-s) - \delta (t-s) \big) \equiv 0$. The Araki-Woods construction (\ref{eq:AW}) is then an integrated version of (\ref{eq:AWqwn}).

Returning to the non-Fock quantum stochastic model, we see that the thermal noise models are not unitarily equivalent to the vacuum quantum stochastic models, and it would be unphysical to interpret (\ref{eq:AW}) as involving a combination of stimulated and spontaneous noise quanta ($b_-$ and $b_+$ respectively).

%%%%%%%%%%%%%%%%%%%%%%%%%%%%%%%%%%%%%%%%%%%%%%%%%%%%%%%%%%%%%%%%%%%%%
%%%%%%%%%%%%%%%%%%%%%%%%%%%%%%%%%%%%%%%%%%%%%%%%%%%%%%%%%%%%%%%%%%%%%
%%%%%%%%%%%%%%%%%%%%%%%%%%%%%%%%%%%%%%%%%%%%%%%%%%%%%%%%%%%%%%%%%%%%%

\section{Algebraic Formulation of the Unruh Effect}
It is convenient to recast quantum fields in curved spacetimes in the language more familiar to quantum probabilists. For the remainder of this paper, we shall use units where $c=1$ and $\hbar = 1$. We fix a spacetime $(\mathscr{M}, g)$ where $\mathscr{M}$ is a four dimensional pseudo-Riemannian manifold and $g$ is metric with signature $(+,-,-,-)$. We assume the existence of a mapping $\widehat{\phi}$ from the set of complex-valued smooth functions of compact support on $\mathscr{M}$ into a set of operators: we denote by $\mathscr{A}$ the unital *-algebra generated by the $\widehat{\phi} (f)$. The map $\widehat{\phi} $ is the Klein-Gordon quantization functional and should be *-linear. It should also satisfy $\widehat{\phi} \big( ( \square +m^2 ) f \big) =0 $ and the Einstein causality condition $ [ \widehat{\phi} (f_1) , \widehat{\phi} (f_2) ] = i E (f_1 , f_2 ) \, I$ where $E= E^{\mathrm{ret.}} - E^{\mathrm{adv.}}$ is the difference between the retarded and advanced Green's functions.

Given a state $\omega$ on $\mathscr{A}$, one can construct the corresponding GNS representation $(\mathscr{H}_\omega , \pi_\omega , \Omega_\omega )$ of $\mathscr{A}$. In the case where $\mathscr{M}$ is not compact, then the GNS representations for different $\omega$ will typically be unitarily inequivalent.

Next, let $v$ be a complete tangent vector field on $\mathscr{M}$ and let $\{ e^{tv} : t \in \mathbb{R} \}$ be the one-parameter group of diffeomorphisms it generates. A family of automorphisms $\alpha_t$ is then generated on $\mathscr{A}$ according to
\begin{eqnarray}
\alpha_t \bigg( \widehat{\phi} (f_1 ) \cdots \widehat{\phi} (f_n) \bigg)
= \widehat{\phi} (f_1 \circ e^{-tv} ) \cdots \widehat{\phi} (f_n \circ e^{-tv} ) .
\end{eqnarray}
We shall be interested in the case where $v$ is a time-like Killing vector field, in which case the diffeomorphisms $e^{tv}$ are isometries on the manifold. 

We recall the Kubo-Martin-Schwinger (KMS) boundary condition for a state $\omega$, a one-parameter group of isometries $\alpha_t$, and a parameter $\beta >0$: set $F_{A,B}(t_1,t_2 ) = \omega ( \alpha_{t_1} ( A )\alpha_{t_2} (B) ) $ then we require $F_{A,B} $ to have an analytic continuation into the region $\{ (z_1 , z_2 )  \in \mathbb{C}^2: 0 \le \text{Im} \, (z _2 - z_1 ) \le \beta \}$, that it be bounded and continuous at the boundary of the region, and that
\begin{eqnarray}
   F_{A,B} (t_1, t_2+i \beta ) = F_{B,A} (t_2 , t_1 ).
	\end{eqnarray}
Here $\beta >0$ is the inverse temperature and $\omega$ is said to be a thermal state.

In order to describe a uniformly accelerated observer in Minkowski spacetime with proper acceleration $a>0$, it is convenient to introduce \textit{radar coordinates} $(\eta , \xi , y ,z)$ given by
\begin{eqnarray}
    t= \frac{1}{a} e^{a \xi} \sinh (a \eta ), \quad x = \frac{1}{a}e^{a \xi} \cosh (a \eta)
\end{eqnarray}
and one then sees that $dt \, dx = e^{2a \xi} \,  d\eta \, d\xi$ and $dT^2 = e^{2a \xi} (d\eta^2 - d\xi^2) - dy^2 -dz^2$. The curve $\xi ,y,z \equiv 0$ describes the world line of a particle accelerating with a constant proper acceleration $a$. 

Varying the radar coordinates over all real values leads to the region $\mathbb{W}$ given $\{(t,x) : 0< |t| <x \}$ called the \textit{right Rindler wedge}.

\begin{figure}[h]
    \centering
    \includegraphics[width=0.40\linewidth]{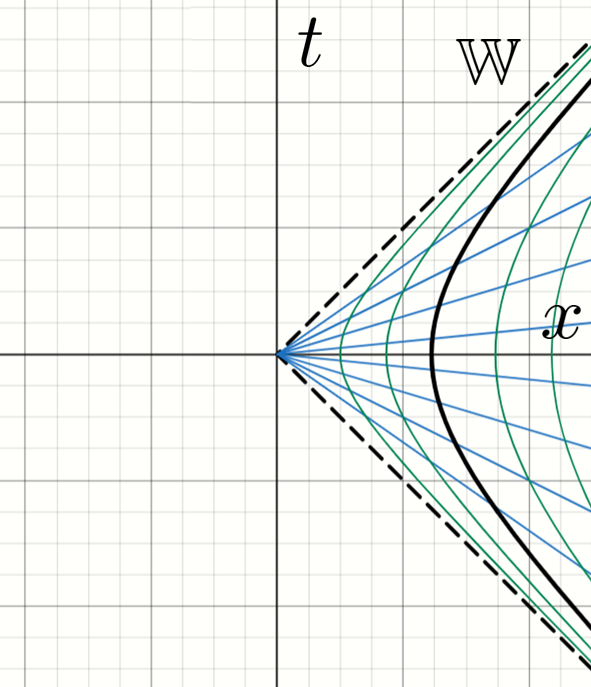}
    \caption{The Rindler wedge $\mathbb{W}$. Curves of constant $\tau$ (blue) and constant $\xi$ (green) are plotted. The special case $\xi=0$ gives the worldline of a particle accelerating at fixed proper-acceleration $a>0$ (bold). }
    \label{fig:enter-label}
\end{figure}

The vector field $ \frac{\partial }{\partial \eta} = a (x \frac{\partial }{\partial t} -  t \frac{\partial }{\partial x} )$ is a time-like Killing vector field in $\mathbb{W}$. Following Sewell \cite{Sewell}, the Unruh effect may now be restated in algebraic terms - specifically as a special case of a Theorem of Bisognano and Wichmann \cite{BisWich}. Let $\mathscr{A}_M$ and $\mathscr{A}_R$ be the algebras generated by the field observables with test functions in Minkowski space and the Rindler wedge, respectively.
Then the restriction $\omega_R$ of the vacuum state of $\mathscr{A}_M$ to $\mathscr{A}_R$ is an invariant under the automorphism group generated by $\frac{\partial }{\partial \eta}$ and moreover satisfies the KMS condition for this group with inverse temperature $\beta = \frac{2 \pi}{a}$.

\subsection{Uniformly Accelerating Two-Level Detector}
We shall adapt the well known simple model, see for instance the presentation in \cite{Alves} and references therein.
We now model a simple detector moving in spacetime. The detector is taken to move along a world-line $x (\tau)$ parametrized by its proper time $\tau$ though we are interested only in its internal quantum degrees of freedom and here we take it just to be a two-level atom with internal frequency $\Omega$ as measured in its instantaneous rest frame. We take $| g \rangle$ and $| e \rangle$ to be the ground and excited state of the detector and introduce the lowering operator $\sigma_- = |g \rangle \langle e |$ along with the raising operator $\sigma_+ = \sigma_-^\ast$.

The detector just feels the quantum field at its current spacetime location and so we revert to the Dirac picture where we subtract off the free dynamics of both the detector itself ($H_{\text{det.}} = \frac{1}{2} \Omega \sigma_z$) and the field (ultimately described by a family of automorphisms generated by a congruence of integral curves that includes the world-line of the detector). The unitary family of relevance to the detector and the local field is taken to be of the form
$d U (\tau) = - i \Upsilon (\tau ) \, U (\tau) $ initialized as $U(0) = I_{\text{det.}} \otimes I_{\text{field}}$ with minimal coupling interaction
\begin{eqnarray}
 -i \Upsilon (\tau ) = \big( \sigma_- e^{-i \Omega \tau } - \sigma_+ e^{i \Omega \tau } \big) \otimes  \widehat{\phi } \big( x(\tau ) \big)
\end{eqnarray}
Here the system coupling is through $\sigma_y = i (\sigma_- - \sigma_+)$ where the lowering and raising terms pick up counter-rotating phases due to $H_{\text{det.}}$. Likewise, we just have the contribution of the field at $x (\tau )$.

The detector is assumed to be initially in its ground state with the field in the (Minkowski) vacuum $\langle \cdot \rangle$. 
We note the two point function for the field in the Minkowski vacuum state is
\begin{eqnarray}
\langle \widehat{\phi } (x_1 ) \widehat{\phi } (x_2)  \rangle = \lim_{\epsilon \to 0^+} \frac{1}{ 4 \pi^2}
\frac{1}{( x_1-x_2 - i \epsilon \hat t )^2}
\end{eqnarray}
where $(x)^2 \equiv \eta_{ij } x^i x^j$, $\hat t$ is any future pointing time-like vector, and the limit is understood in the distributional sense. One may show that for the fields along the worldline of the uniform accelerating observer
\begin{eqnarray}
\langle \widehat{\phi } \big( x ( \tau_1) \big) \widehat{\phi }\big( x( \tau_2) \big) \rangle = - \lim_{\epsilon \to 0^+} \frac{a^2}{ 4 \pi^2}
\frac{1}{4 \sinh \left( \frac{ a (\tau_1 - \tau_2 -i \epsilon )}{2} \right) }.
\end{eqnarray}

The objective is to calculate the probability rate for the detector to transition to its excited state, which should require an absorption of a quantum from the field. Clearly, if the detector moves as an inertial body, this rate should be zero. This changes if the detector is uniformly accelerated and to this end we take its world-line to be $( \frac{1}{a} \sinh (a \tau ) , \frac{1}{a} \cosh (a \tau ) , 0,0)$ in standard coordinates: this corresponds to $\eta \equiv \tau$ and $\xi = y=z =0 $ in radar coordinates. The parameter $\tau$ will be the proper time along the world-line.

\subsection{Quantum Stochastic Limit for a Uniformly Accelerating Detector}

The standard argument at this point is to perform a Fermi Golden Rule calculation, however, we shall frame this as an asymptotic quantum stochastic limit. We introduce a coupling constant $\lambda$ and rescale proper time as $ \tau \to \tau /\lambda^2$ (the van Hove rescaling). Then we take $\lambda \to 0^+$. Note that this is a rescaling of the proper time and here we have the replacement $t \to t = \frac{1}{a} \sinh (a \tau / \lambda^2 ) \approx \frac{1}{a} e^{a \tau / \lambda^2 }$ in the fixed inertial frame for $\tau >0$.

With this rescaling, we define $ U_\lambda (\tau ) \equiv U( \tau / \lambda^2 )$ so that $d U _\lambda (\tau) = - i \Upsilon _\lambda (\tau ) \, U (\tau) $ with 
\begin{eqnarray}
 -i \Upsilon _\lambda (\tau ) = \frac{1}{\lambda} \big( \sigma_- e^{-i \Omega \tau / \lambda^2} - \sigma_+ e^{i \Omega \tau / \lambda^2} \big) \otimes  \widehat{\phi } \big( x(\tau / \lambda^2) \big) .
\end{eqnarray}
From this, we obtain the Dyson series expansion
\begin{eqnarray}
 U_\lambda (\tau ) = I_{\text{det.}} \otimes I_{\text{field}} + \sum_{n \ge 1} (-i)^n \int_{\Delta_n (\tau )} d\tau_n \cdots d \tau _1 \,
\Upsilon _\lambda (\tau_n ) \cdots \Upsilon _\lambda (\tau_1 )
\end{eqnarray}
where $\Delta_n (\tau )$ is the simplex $\{ (\tau_n , \cdots , \tau_1 ): t \ge \tau_n > \cdots \ge \tau_1 \ge 0\}$.

The idea of the quantum stochastic limit \cite{AFLI,AFLII,ALV} rests on the observation that if $g$ is an integrable function with $\gamma = \int_{-\infty} ^\infty g(t) \, dt$ then we 
\begin{eqnarray}
\lim_{ \lambda \to 0^+} \frac{1}{\lambda^2} \, g ( \frac{\tau }{\lambda^2} ) = \gamma \, \delta (\tau) .
\end{eqnarray}
To see this, we note for any Schwartz test function $f$ that $\int_{-\infty} ^\infty f(\tau )\frac{1}{\lambda^2} g ( \frac{\tau }{\lambda^2} )  \, d\tau =\int_{-\infty} ^\infty f(\lambda^2 t )  g (t) \, dt \to \gamma \, f(0)$. Under the correct rescaling, the fields behave like delta-correlated quantum white noises. In our case, we should focus on the rescaled operators $\frac{1}{\lambda}  e^{\pm i \Omega \tau / \lambda^2}  \widehat{\phi } \big( x(\tau / \lambda^2) \big) $.

We consider the rescaled two-point function 
\begin{eqnarray}
\frac{1}{\lambda^2} \, e^{- i \Omega (\tau_1 - \tau_2)}
\langle \widehat{\phi } \big( x ( \tau_1) \big) \widehat{\phi }\big( x( \tau_2) \big) \rangle \to \gamma \, (\tau_1 - \tau_2 )
\end{eqnarray}
with 
\begin{eqnarray}
\gamma = - \lim_{\epsilon \to 0^+} \int_{-\infty}^\infty \frac{a^2}{ 4 \pi^2}
\frac{e^{- i \Omega \tau } \, d \tau }{4 \sinh \left( \frac{ a (\tau_1 - \tau_2 -i \epsilon )}{2} \right) }
\equiv \frac{\Omega}{4 \pi} \, n (\Omega ) 
\end{eqnarray}
where $ n (\Omega ) = \frac{1}{ e^{2 \pi \Omega /a} - 1}$ and the integral may be evaluated by the residue Theorem.

We therefore have the following asymptotic limits
\begin{eqnarray}
\frac{1}{\lambda} e^{- i \Omega \tau / \lambda^2}  \,   \widehat{\phi } \big( x(\tau / \lambda^2) \big) \to \sqrt{\frac{\Omega}{4 \pi}  } 
 \,  b (\tau )^\ast  , \nonumber \\
\frac{1}{\lambda}  \,  e^{+ i \Omega \tau / \lambda^2}  \widehat{\phi } \big( x(\tau / \lambda^2) \big) \to \sqrt{\frac{\Omega}{4 \pi}  }   \,
b (\tau )  .
\end{eqnarray}
where the limit processes are (gaussian) thermal quantum white noises satisfying
\begin{eqnarray}
[ b (\tau_1 ) , b(\tau_2 )^\ast ] = \delta ( \tau_1 - \tau _2 ) ,
\end{eqnarray}
with
\begin{eqnarray}
\langle b(\tau_1) b (\tau_2 )^\ast \rangle = \big( n (\omega ) + 1 \big) \, \delta (\tau_1 - \tau _2 ) , \quad
\langle b(\tau_1)^\ast b (\tau_2 ) \rangle =  n (\omega )  \, \delta (\tau_1 - \tau _2 ) ,
\end{eqnarray}
along with $ \langle b(\tau_1) b (\tau_2 ) \rangle = 0 = \langle b(\tau_1)^\ast b (\tau_2 )^\ast \rangle $.

More rigorously, we look at the integrated versions of these quantum white noises, formally $B( \tau ) = \int_0^\tau b(\sigma ) \, d \sigma$,  and these are non-Fock quantum Wiener processes with quantum Ito table
\begin{eqnarray}
dB d B^\ast = \big( n (\Omega )+1 \big) \, dt , \quad
dB^\ast d B =  n (\Omega )  \, dt \quad dBdB=dB^\ast dB^\ast = 0.
\end{eqnarray}

The limit evolution operator is then given by the quantum stochastic process satisfying the formal quantum white noise differential equation
\begin{eqnarray}
\frac{d}{dt} U (t) = \sqrt{ \frac{\Omega }{4 \pi}} \bigg( \sigma_- \otimes b(\tau )^\ast  - \sigma_+ \otimes b (\tau ) \bigg) U(t) .
\end{eqnarray}
and this corresponds to the non-Fock quantum stochastic differential equation \cite{HL85}
\begin{eqnarray}
d U(\tau ) &=&  \bigg( \sqrt{  \frac{\Omega }{4 \pi} } \, \sigma_- \otimes dB(\tau )^\ast  - \sqrt{ \frac{\Omega }{4 \pi} } \, \sigma_+ \otimes dB (\tau ) \nonumber \\
&& - \frac{1}{2}  \frac{\Omega}{4 \pi} \big( n (\Omega )+1 \big) \sigma_+ \sigma_- \otimes d \tau -  
\frac{1}{2} \frac{\Omega }{4 \pi} n (\Omega )  \sigma_- \sigma_+ \otimes d \tau \bigg) U(t) .
\end{eqnarray}
Note that we may represent the non-Fock processes on the doubled up Fock space with $B(\tau) \equiv \sqrt{n(\Omega ) + 1} A_1 (\tau ) \otimes I_2 + \sqrt{n(\Omega )} I_1 \otimes A_2 (\tau )^ast$ with the joint vacuum state as cyclic state.

This describes the detector as an open system with Lindblad generator
\begin{eqnarray}
\mathcal{L} (X) = \frac{\Omega }{4 \pi} \big( n (\Omega )+1 \big) \mathcal{L}_{\sigma_+ \sigma_-} (X) +  
\frac{\Omega }{4 \pi}  n (\Omega ) \mathcal{L}_{ \sigma_- \sigma_+ } (X) ,
\end{eqnarray}
where $\mathcal{L}_L (X) =\frac{1}{2} [L^\ast , X ]L + \frac{1}{2} L^\ast [X,L]$.

The state of the detector may be described by a $2 \times 2$ density matrix $\rho (\tau)$ where $ \text{tr} ( \rho(\tau ) \, X ) = \langle \text{tr} ( \rho \, U(\tau )^\ast (X \otimes I ) U(\tau ) \rangle$ for any detector observable $X$. From this we obtain the master equation $\frac{d}{d \tau}  \rho (\tau ) = \mathcal{L}^\star \rho (\tau)$ where $\mathcal{L}^\star$ is the Liouvillian defined as the adjoint of the Lindbladian though the duality $\text{tr} ( \mathcal{L}^\star \rho \, X ) = \text{tr} ( \rho \, \mathcal{L} (X) ) $. 

It is easy to see that $ \rho (\tau )$ will relax to the thermal state $\rho_{\text{thermal}} =  n_F (\Omega ) \sigma_- \sigma_+ +  \big( 1- n_F (\Omega ) \big)\sigma_+ \sigma_-$. Here $n_F (\Omega ) = \frac{n(\Omega) +1}{2n(\Omega ) +1} = \frac{1}{ e^{ 2 \pi \Omega /a}+1 }$ is the \textit{fermion} average occupation number at thermal equilibrium (as expected for a two level atom consisting of a ground and excited state). In fact, we solve the master equation exactly to get
\begin{eqnarray}
\langle e| \rho (\tau ) | e \rangle &=& \frac{n (\Omega ) }{2 n (\Omega ) +1 } + \left(\langle e| \rho (\tau ) | e \rangle - \frac{n (\Omega ) }{2 n (\Omega ) +1 } \right) e^{- (2 n (\Omega ) + 1) \Omega \tau / 4 \pi } , \nonumber \\
\langle g| \rho (\tau ) | g \rangle &=& \frac{n (\Omega ) +1 }{2 n (\Omega ) +1 } + \left(\langle g| \rho (\tau ) | g \rangle - \frac{n (\Omega )+1 }{2 n (\Omega ) +1 } \right) e^{- (2 n (\Omega ) + 1) \Omega \tau / 4 \pi } , \nonumber \\
\langle e| \rho (\tau ) | g \rangle &=&   \langle e| \rho (\tau ) | g \rangle  e^{- (2 n (\Omega ) + 1) \Omega \tau / 8 \pi } .
\end{eqnarray}

A gauge process $\Lambda (\tau )$ cannot be defined for non-Fock quantum stochastic calculus. 

%%%%%%%%%%%%%%%%%%%%%%%%%%%%%%%%%%%%%%%%%%%%%%%%%%%%%%%%%%%%%%%%%%%%%
%%%%%%%%%%%%%%%%%%%%%%%%%%%%%%%%%%%%%%%%%%%%%%%%%%%%%%%%%%%%%%%%%%%%%
\section{Discussion and Conclusion}
We have re-derived the transformations of quantum stochastic evolutions (\ref{eq:FR_trans}) given in \cite{Frigerio_Ruzzier}, but in a more direct manner. In their treatment, they consider 1+1 dimensional spacetime and have two bosonic fields: one traveling from the right and one from the left. The multiple field case does not pose any problem as the same transformations (\ref{eq:trans_gamma}) and (\ref{eq:FR_trans}) apply to all scattering, annihilation, creation, and time processes.

 In principle, we could even handle the usual 1+3 spacetime by introducing a bosonic noise for each unit vector in 3-space. This would require us to take the multiplicity space for the noise to be infinite-dimensional, but this is inherent in the Hudson and Parthasarathy formalism \cite{HP84}.

The transformation rules for the quantum stochastic calculus apply to systems that moving as inertial systems as described by inertial observers. If the system is slowly accelerating then we may expect this to be still approximately true, however, this cannot be the case in general as illustrated by the Unruh effect for uniformly accelerated systems. The Unruh effect essentially only says that a uniformly accelerated observer finds that the quantized Klein-Gordon field in the Minkowski vacuum state acts to thermalize the observer at the Unruh temperature (proportional to the proper acceleration) - it does not strictly speaking make any statement about the existence of radiation detectable as particles - the non-existence of the gauge process $\Lambda (\tau )$ is consistent with the claim that there are no Unruh particles as such.

\end{document}